# Polarization-induced Rashba spin-orbit coupling in structurally symmetric III-Nitride quantum wells

V.I. Litvinov[*]

*WaveBand/ Sierra Nevada Corporation,*
*15245 Alton Parkway, Suite 100, Irvine, CA 92618*

**Abstract.** The effective linear coupling coefficient and the total spin-splitting are calculated in Ga- and N- face InGaN quantum wells. Alloy content, geometry, and gate voltage affect an internal field and an electron density distribution in the growth direction that has direct effect on a spin-splitting. The sign of structural inversion asymmetry (SIA) spin-orbit coupling coefficient depends on an internal electric field in the well that results in different signs for Ga-face and N-face III-Nitride structures. The effective linear coupling coefficient is always positive because of the Dresselhaus-type contribution that is a major one in quantum wells under consideration. The magnitude of the spin-splitting is comparable with that experimentally observed in III-Nitrides and III-V zinc-blende structures.

---

[*] E-mail address: vladimir.litvinov@sncorp.com

# 1. Introduction.

Gate-voltage manipulation of electron spins in semiconductor heterostructures is the subject of intensive study in regard with the emerging semiconductor spintronic devices such as spin transistors, spin light-emitting diodes, and quantum computers.[1,2] Electrical manipulation of spins is possible via the momentum-dependent electron spin-splitting in gated semiconductor heterostructures. The splitting is a result of a spin-orbit interaction and a bulk spatial inversion asymmetry (BIA)[3], and (or) a structural inversion asymmetry caused by interfaces (SIA or Rashba term)[4].

Among the perspective spintronic materials, the structures made of GaN/AlN/InN are known as material systems that may deliver room-temperature ferromagnetism (see, for instance, review[5]). Recently, it was shown that the wide bandgap wurtzite GaN-based materials are suitable for using in gate-modulated spintronics: the Rashba spin splitting in a AlGaN/GaN modulation doped channel was found to be about 1 meV [6] that is comparable to that in cubic GaAs-based layers.

The spin-splitting in AlGaN/GaN heterostructures was found to be larger than 5 meV. [7,8] This spin-splitting was explained in Ref.[9] by zone-folding effects. Zone-folding in a wurtzite (W)-structure results in the spin-orbit interaction between close conduction bands. The zone-folding mechanism differs from the usual BIA and SIA ones where the electron spin-orbit interaction stems from the coupling between conduction and valence bands and become smaller when the bandgap increases. On the other hand, since the experimental data were obtained from the beatings of Shubnikov-de Haas oscillations, the results have recently been questioned as the oscillation beatings might alternatively be explained by an intersubband scattering in a magnetic field rather than a large spin-splitting.[10] Irrespective to that which interpretation is correct, it is clear that the experimental data should be analyzed taking into account all contributions: linear SIA term[6] as well as BIA $k^3$- and $k$-linear terms, $\hbar \vec{k}$ being the electron momentum. The linear BIA spin-orbit term[11,12] is specific to a W-bulk crystal and absent in a zinc-blende (ZB) one.

SIA and BIA contributions to the spin-splitting in ZB III-V structures are pretty much known.[1] Since one expects smaller Rashba spin splitting in the larger bandgap materials, the nitride structures have attracted much less attention than low bandgap ZB materials in regards with spintronic applications. However, recent experimental data on a bandgap in InN (0.69 eV)[13] might introduce InGaN alloys into a spintronic material family. This paper addresses the spintronic properties of InGaN/GaN QW and discusses how various spin-orbit terms contribute to the total zero-field spin-splitting.

Comparing the BIA and SIA in ZB and W-materials it is important to realize the essential difference between the two. In a symmetric and unbiased ZB [100]-oriented quantum well (QW) the SIA does not exist and the Rashba coupling turns to zero. On the contrary, even structurally symmetric and unbiased W-QW is electrically asymmetric due to built-in polarization fields. This results in non-zero Rashba coupling coefficient, which depends on internal fields. Another difference is that the c-axis oriented two-dimensional (2D) electrons in a W-structure feel the spin-orbit interaction in a different way compared to a ZB structure. It is known that $\vec{k}$-dependent electron spin configuration in ZB [100]-QW depends on the relation between the competing Dresselhaus and Rashba terms.[14,15,16] In a c-axis oriented W-structure both the Rashba and the Dresselhaus-type terms induce the same electron spin configuration. This means that both $k$-linear and $k^3$-terms contribute to an effective linear spin-orbit coupling.

In this paper, the calculations of spin-splitting for a symmetric nitride QW are performed for the first time. The general equation for the Rashba coefficient in nitride QW has been derived previously[6]. However, this equation itself does not speak for the polarization fields and wave functions one has to know in order to calculate the Rashba coefficient in a QW. In Ref.[6] the calculations were performed all the way down for a modulation-doped heterostructure. In the modulation-doped structure the Rashba splitting exists irrespective to that whether the polarization field is present or not, so the difference between GaAs and GaN is in numbers only. In this paper the calculations are done for a symmetric QW where the zero-field spin-splitting exists only in GaN-based and is absent in [100] ZB structures.

It is found how the magnitude of the effective coupling depends on internal fields, resulting in Rashba coefficients of opposite signs in Ga- and N-face QWs. The spin splitting is calculated as a function of gate bias and QW width. The Rashba coefficient may change its sign under appropriate gate voltage. The Dresselhaus-type term constitutes the major contribution to an effective linear coupling and the overall spin splitting in QWs.

**2. Effective spin-orbit interaction**

In a bulk W-crystal the spin-orbit interaction comprises the invariants of the $C_{6v}$ point-group including linear and cubic BIA terms[17]:

$$H_{WBIA} = \lambda (\vec{\sigma} \times \vec{k}) \vec{z} + \frac{1}{2} \lambda_l \left[ k_z^2 (\vec{\sigma} \times \vec{k})_z + (\vec{\sigma} \times \vec{k})_z k_z^2 \right] + \frac{1}{2} \lambda_t \left[ k_\parallel^2 (\vec{\sigma} \times \vec{k})_z + (\vec{\sigma} \times \vec{k})_z k_\parallel^2 \right], \quad (1)$$

where $\vec{z}$ is the c-axis in wurtzite crystal, $k_\parallel^2 = k_x^2 + k_y^2$, $\vec{\sigma}$ are Pauli matrices, the coefficients $\lambda$ and $\lambda_l, \lambda_t$ describe the magnitude of the BIA spin-orbit $k$-linear and $k^3$-terms, respectively.

In a structure confined in c-direction, $k_z$ is quantized and the effective linear spin-orbit coupling Eq.(1) takes the form:

$$H_{BIA} = \alpha_{BIA} (\sigma_x k_y - k_x \sigma_y), \quad \alpha_{BIA} = \lambda + \lambda_l < k_z^2 >, \quad (2)$$

where $<...>$ means the average with the QW envelope wave function $\Phi(z)$ that belongs to the energy level $\varepsilon_1$.

The effective BIA spin-orbit interaction Eq.(2) has exactly the same form as the Rashba term $H_R = \alpha_R (\sigma_x k_y - k_x \sigma_y)$ does, thus the total spin-orbit coupling Hamiltonian in W-structures is given as an effective Rashba coupling:

$$H_{SO} = H_{BIA} + H_R = \alpha_{eff} (\sigma_x k_y - k_x \sigma_y), \quad \alpha_{eff} = \alpha_R + \alpha_{BIA}. \quad (3)$$

The eigenspinor and electron spin-splitting resulting from the Hamiltonian Eq(3) follow:

$$u_\pm = \frac{1}{\sqrt{2}} \begin{pmatrix} 1 \\ \mp i e^{-i\varphi} \end{pmatrix}, \quad tan\,\varphi = k_y / k_x, \quad \Delta\varepsilon = 2\alpha_{eff} k_\parallel. \quad (4)$$

For comparison, the spin-orbit interaction in ZB-QW is given as

$$H_{SO} = H_D + H_R = \gamma_D (\sigma_x k_x - k_y \sigma_y) + \gamma_R (\sigma_x k_y - k_x \sigma_y), \quad (5)$$

where $\gamma_D$ and $\gamma_R$ are the Dresselhaus and Rashba coefficients, respectively. The eigenspinors and the spin splitting resulting from the Hamiltonian Eq.(5) are given as

$$u_{\pm} = \frac{1}{\sqrt{2}}\begin{pmatrix} 1 \\ \mp e^{-i\theta} \end{pmatrix}, \quad tan\,\theta = \frac{\gamma_D \sin\varphi + \gamma_R \cos\varphi}{\gamma_R \sin\varphi + \gamma_D \cos\varphi},$$

(6)

$$\Delta\varepsilon = 2\gamma k_{\parallel}, \quad \gamma = \sqrt{\gamma_D^2 + \gamma_R^2 + 2\gamma_D\gamma_R \sin 2\varphi}.$$

As it follows from Eq.(6), the conditions $\gamma_D = \pm\gamma_R$ make the eigenspinors Eq.(6) independent on an electron momentum that results in the suppression of a spin relaxation as discussed in Refs. [16].

The $\vec{k}$-dependent spin orientation of in-plane electrons in W-structures is determined by an effective Rashba coupling Eq.(3) that accounts for all three contributions: linear BIA, cubic BIA, and linear SIA, likewise in an [111]-oriented ZB-QW.[18] Providing all of them of the same sign, it might increase the observable linear spin-splitting as compared to [100] oriented ZB-QW where BIA and SIA may compensate each other in the magnitude of $\gamma$ Eq.(6).

The gate-voltage and structure-specific effective spin-orbit coupling coefficient $\alpha_{eff}$ will be calculated below for structurally-symmetric InGaN/GaN QW.

### 3. Rashba coefficient in an Ga-face QW.

The general expression for the Rashba coefficient in W-structures has been obtained in Ref. [6]. It includes contributions from the left (LB), right (RB) barriers and the well (WL) as given below:

$$\alpha_R = P_1 P_2 \left[ \Phi^2(0)(\beta_{WL} - \beta_{LB}) - \Phi^2(LB)(\beta_{WL} - \beta_{RB}) + \left\langle \frac{\partial\beta}{\partial z} \right\rangle_{LB} + \left\langle \frac{\partial\beta}{\partial z} \right\rangle_{WL} + \left\langle \frac{\partial\beta}{\partial z} \right\rangle_{RB} \right]$$

$$\beta(z) = \frac{\Delta_3}{(G-\varepsilon_1)(\Lambda-\varepsilon_1) - 2\Delta_3^2},$$

(7)

where $P_1, P_2$ are momentum matrix elements, $\Lambda = E_v^0 + S_1 + \tilde{V}$, $G = E_v^0 + \Delta_1 - \Delta_2 + S_1 + S_2 + \tilde{V}$, $S_1 = D_1\varepsilon_{zz} + D_2(\varepsilon_{xx} + \varepsilon_{yy})$, $S_2 = D_3\varepsilon_{zz} + D_4(\varepsilon_{xx} + \varepsilon_{yy})$, $\varepsilon_{ij}$ are the strain components, $D_i$ are the deformation potentials, $\Delta_1$ and $\Delta_{2,3}$ are parameters of the crystal field and spin-orbit interaction, respectively, $E_v^0$ is the valence band edge position before the strain, crystal-field and spin-orbit splitting taken into account. We assume that z-dependent valence and conduction band positions account for the band offsets and $\tilde{V}$ includes contributions from both an external electric field (bias) and a self-consistent potential in an inhomogeneous structure.

Inspection of Eq.(7) shows that in an unbiased structurally symmetric QW without internal electric fields the Rashba coefficient is equal to zero. The internal electric fields that are specific to GaN-based QWs determine the magnitude of $\alpha_R$ in a structurally-symmetric $GaN/In_xGa_{1-x}N$ QW. Electric fields in a three-layer system of thickness $d = d_{LB} + d_{WL} + d_{RB}$ biased with the external voltage $V$ can be found from the equations that follow:

$$\begin{cases} \varepsilon_0\varepsilon_{LB}E_{LB} + P_{LB} = \varepsilon_0\varepsilon_{WL}E_{WL} + P_{WL} \\ \varepsilon_0\varepsilon_{RB}E_{RB} + P_{RB} = \varepsilon_0\varepsilon_{WL}E_{WL} + P_{WL} \\ E_{LB}d_{LB} + E_{RB}d_{RB} + E_{WL}d_{WL} = -V \end{cases} \tag{8}$$

where $P_i(x) = P_{sp}(x) + P_{pz}(x)$ is the total polarization that comprises spontaneous and piezoelectric parts, $E_i, \varepsilon_i (\varepsilon_0)$ are the electric field and dielectric permittivity in layer $i$ (vacuum), respectively.

The piezoelectric polarization is present in the compressed $In_xGa_{1-x}N$ layer and equal to zero in the GaN barriers which are assumed to be lattice-matched to the GaN buffer. Material parameters used in numerical calculations are given in Table 1.

Table 1. Parameters of $Ga_{1-x}In_xN$ material system.

| | |
|---|---|
| Effective mass ($m_0$) | $0.2 - 0.09\,x$ |
| Permittivity ($\varepsilon_0$) | $10.28 + 4.33\,x$ |
| Elastic constants ($GPa$) | $C_{13} = 103 - 11\,x\,;\,C_{33} = 405 - 181\,x$ [a] |
| Polarization $\left(\dfrac{C}{m^2}\right)$ | $P_{sp}(x) = -0.034 + 0.029\,x - 0.037\,x^2$ [a] <br> $P_{pz}(x) = 0.106\,x + 0.042\,x^2$ |
| Band gap in relaxed GaInN(eV) | $E_g(x) = 0.69\,x + 3.5(1-x)$ [b] |
| Valence band offset, eV | $\Delta E_v = 0.26\,x$ [c] |
| Spin-orbit split energy (meV), | $\Delta_2 = 6.5 - 6.0\,x$ [c] <br> $\Delta_3 = 13.5 - 9.8\,x$ |
| Crystal-field split energy (meV) | $\Delta_1 = 21$ [c] |
| Interband momentum-matrix elements <br> $P_{1,2} = \hbar\sqrt{E_{1,2}/2m_0}$ | $E_1 = E_2 = 20.0\,eV$ |

a) Reference [19], b) Reference [13], c) Reference [20]

The conduction band offset is calculated using the valence band offset $\Delta E_v$ (Table 1) and the model solid approach that accounts for the alloy and strain effects.[21] The example of the potential profile and the corresponding wave function in Ga-face $GaN/In_xGa_{1-x}N/GaN$ QW are shown in Fig. 1. The z-axis is pointed from the substrate to the growth direction.

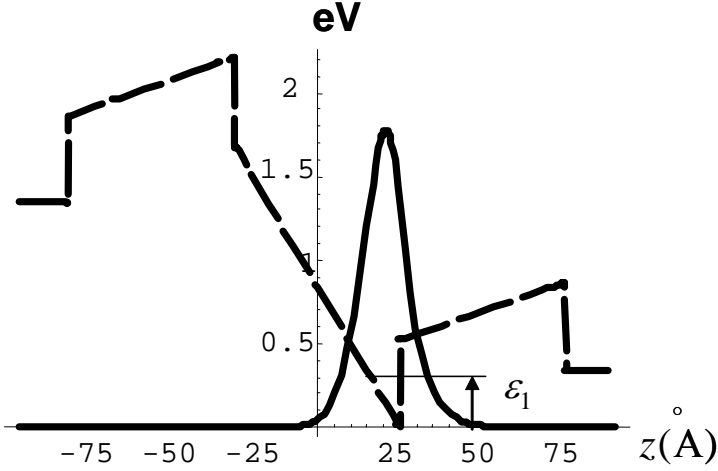

Fig.1. *Potential profile (eV, dashed) and wavefunction $\Phi^2(z)$ (a.u., solid) in $50\,\overset{\circ}{A}/40\,\overset{\circ}{A}/50\,\overset{\circ}{A}$ QW, $x=0.5$, $\varepsilon_1 = 0.37\,eV$, $V=1\,V$.*

The conduction band profile shown in Fig.1, has been verified with the self-consistent Poisson-Schrodinger solver. The deviations from the shape of Fig.1 are negligible even in AlGaN/GaN heterostructure. In InGaN/GaN quantum well the electrons do not form a two-dimensional electron gas near the InGaN/GaN interface (polarization charge is negative). The situation is different from that in the GaN well where electrons accumulate near GaN/AlGaN interface (polarization charge is positive).

Linear $\lambda$ and cubic $\lambda_l, \lambda_t$ BIA coupling coefficients in III-Nitrides are unknown. The Dresselhaus constant in GaAs was estimated as $\gamma_D \approx 24\,eV\,\overset{\circ}{A}^3$.[22] Since in GaN the spin-orbit interaction is weaker, it is assumed that cubic coupling coefficients in GaN is 10 times less than in GaAs: $\lambda_l = \lambda_t \approx 2.4\,eV\,\overset{\circ}{A}^3$. The linear BIA coefficient was estimated as $\lambda \approx (1.1, 12, 60)\,meV\,\overset{\circ}{A}$ for ZnO, CdS, and CdSe, respectively.[12] In this paper $\lambda = 1.1\,meV\,\overset{\circ}{A}$ is used for numerical calculations. Fig.2 illustrates the well width dependence of the Rashba coefficient Eq.(7) in an unbiased QW. The gate-voltage dependent spin-orbit coupling is shown in Fig.3.

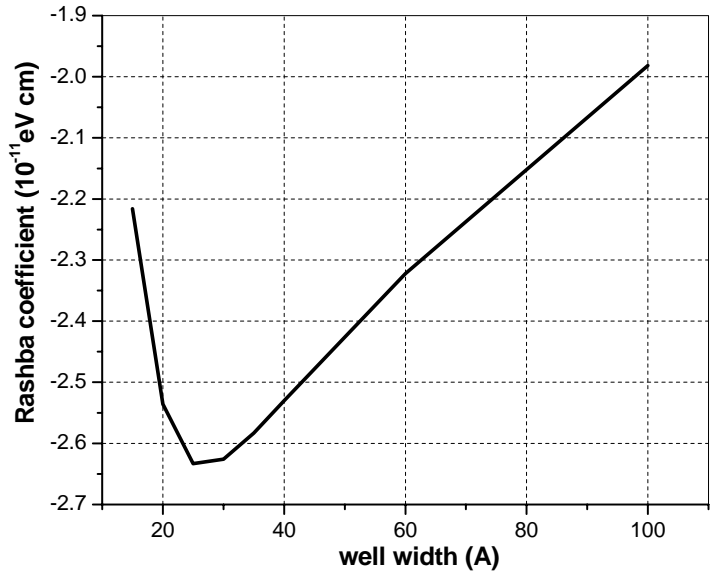

Fig.2. *Rashba coefficient in $GaN/In_{0.5}Ga_{0.5}N/GaN$ QW, $d_{LB} = d_{RB} = 50 \overset{\circ}{A}$.*

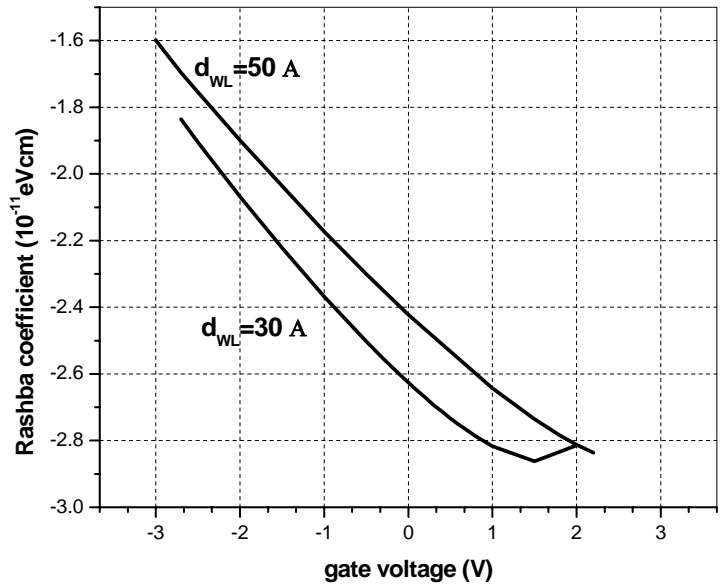

Fig.3. *The external voltage dependence of Rashba coefficient in a $GaN/In_{0.5}Ga_{0.5}N/GaN$ QW: $d_{LB} = d_{RB} = 50\overset{\circ}{A}, d_{WL} = 30\overset{\circ}{A}/50\overset{\circ}{A}$.*

The maximum in absolute value of Rashba coefficient, seen in Figs 1 and 2, stems from the term $\left\langle \frac{\partial \beta}{\partial z} \right\rangle_{WL}$ in Eq.(7) and corresponds to a maximum extent of QW wavefunction localization near the right wall of the QW increasing an electrical asymmetry of the structure (see Fig.1). This extent of asymmetry depends on both the well thickness and the applied voltage. Electron confinement in a QW exists in a certain range of the gate voltage. Outside this range the electric field strongly distorts QW potential profile diminishing the electron confinement. Fig.3. illustrates the actual voltage ranges where the electron confinement exists.

Fig.4 illustrates the Rashba coefficient along with the electric field in the well $E_{WL}$. At $V = -2.7\,\text{V}$ the internal field in the well changes its sign as shown in Fig.4. Starting from the small negative $E_{WL}$ the Rashba coefficient turns positive.

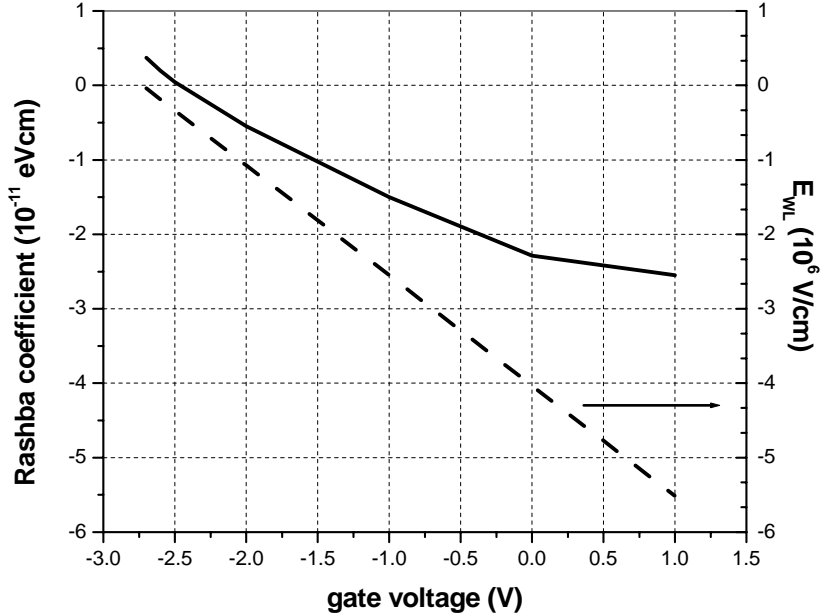

Fig.4. *The external voltage dependence of Rashba coefficient in a $GaN/In_{0.6}Ga_{0.4}N/GaN$ QW: $d_{LB} = d_{RB} = 15\,\text{Å}, d_{WL} = 30\,\text{Å}$ (solid). Dashed is the internal electric field in the well.*

## 4. Rashba coefficient in an N-face QW.

Positive $E_{WL}$ are typical for N-face QWs as illustrated in Fig. 5. The internal electric fields are opposite to those shown in Fig. 1 for Ga-face QWs. The Rashba coefficient for the QW shown in Fig.5 depends on the gate voltage as illustrated in Fig. 6.

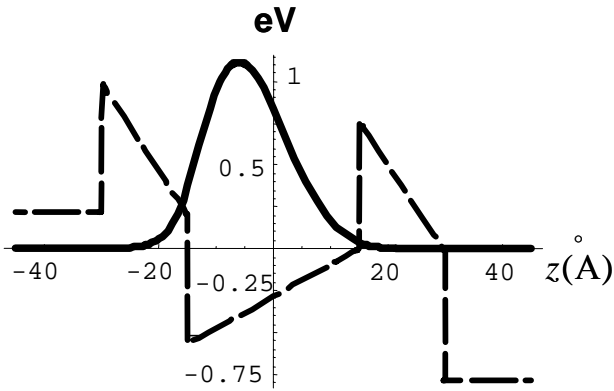

Fig.5. *Potential profile (eV, dashed) and wavefunction $\Phi^2(z)$ (a.u., solid): N-face $15\,\overset{\circ}{A}/30\,\overset{\circ}{A}/15\,\overset{\circ}{A}$ $GaN/In_xGa_{1-x}N/GaN$, $x = 0.5$, $\varepsilon_1 = -0.177\ eV$, $V = 1V$.*

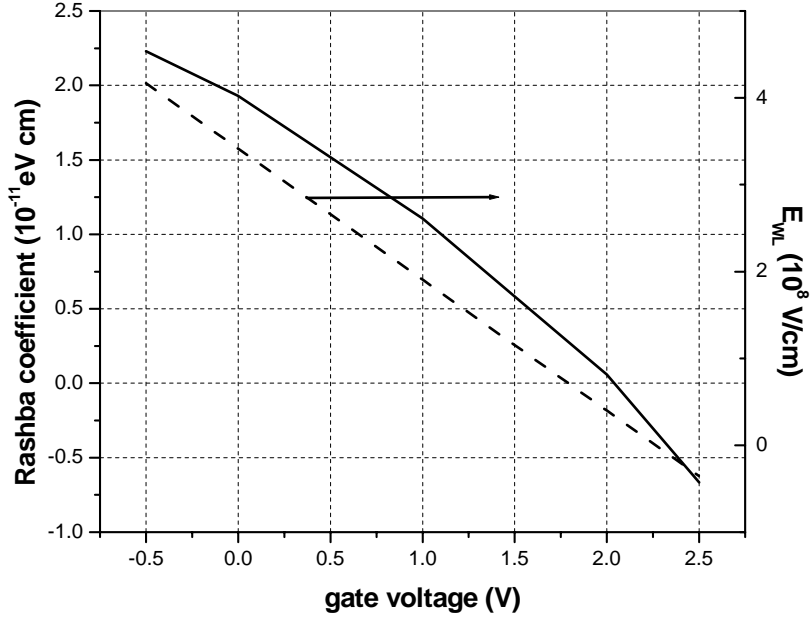

Fig.6. *The external voltage dependence of Rashba coefficient in an N-face GaN/In$_{0.5}$Ga$_{0.5}$N/GaN QW: $d_{LB} = d_{RB} = 15$ Å, $d_{WL} = 30$ Å. Dashed is the internal electric field in the well.*

Similar to that obtained in Ga-face QW (see Fig.4), the Rashba coefficient changes its sign not exactly along with the internal electric field in the well because the barriers biased in their own way also contribute the spin-orbit coupling.

### 5. Spin splitting.

The overall effective linear spin-splitting is determined by an effective Rashba coefficient $a_{eff}$ that depends on BIA contribution $\alpha_{BIA} = \lambda + \lambda_l <k_z^2>$ from Eq.(3). The SIA contribution ($: a_R$) is the one most tunable if the gate voltage varies. However, the BIA term, at least in QWs considered here, is 4-7 times larger than the Rashba term and constitutes the major contribution to an overall spin splitting. Fig. 7 shows the total spin-splitting that includes both SIA and BIA terms in a Ga-face QW. The well width dependence of the spin-splitting is shown in Fig. 8. Spin-splitting in an N-face QW with similar geometry is twice as much less that shown in Fig. 8.

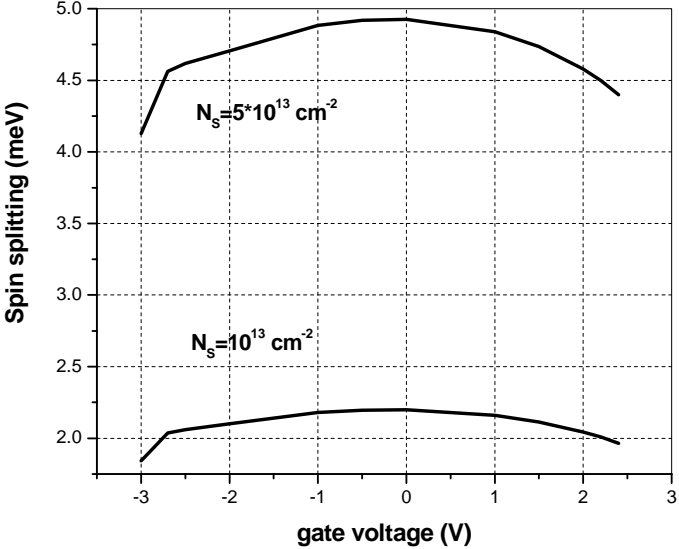

Fig.7. *The external voltage dependence of the total spin splitting at different doping levels $N_S$. Ga-face $GaN/In_{0.5}Ga_{0.5}N/GaN$ QW: $d_{LB} = d_{RB} = 50\,\text{Å}$, $d_{WL} = 50\,\text{Å}$.*

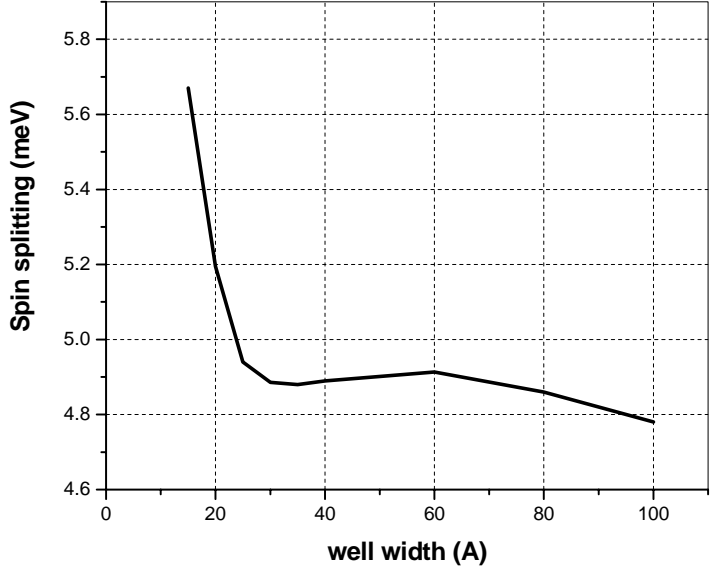

*Fig.8. Total spin splitting in an unbiased Ga-face $GaN/In_{0.5}Ga_{0.5}N/GaN$ QW; $d_{LB} = d_{RB} = 50 \overset{\circ}{\text{A}}$.*

## 6. Conclusions.

In conclusion, the Rashba coefficient itself, the effective linear coupling coefficient, and the overall spin-splitting are calculated in Ga- and N- face InGaN quantum wells. The parameters depend on QW parameters: alloy content, geometry, and gate voltage affect an internal field and an electron density distribution in the growth direction that has direct effect on a spin-splitting.

The SIA spin-orbit coupling coefficients depend on internal electric fields and have different signs for Ga-face and N-face III-Nitride QWs. Since the internal fields are strong, the external voltage-induced field reversal occurs along with the electron tunnel ionization from the well. This narrows the range of the gate voltage within which one could tune the SIA coupling coefficient inverting its sign. The effective linear coupling coefficient is always positive because of the Dresselhaus-type contribution that is a major one in QWs under consideration.

It should be noted that the Rashba contribution to the overall spin-splitting stems from the internal electric fields only since we consider structurally symmetric QWs. However, if the structural asymmetry is also present the Rashba contribution may become the major one. In this case, the ac voltage that periodically changes the sign of Rashba coefficient may deeply modulate the overall spin splitting. It constitutes completely new avenue for engineering the in-plane spin transport in QW.

The magnitude of the spin splitting, estimated in this paper, is comparable with that experimentally observed in III-Nitrides and III-V cubic materials. This suggests that III-nitrides are competitive materials in the perspective spintronics applications.